\documentclass[11pt]{article}

\usepackage[T1]{fontenc}
\usepackage{lmodern}
\usepackage{microtype}
\usepackage{geometry}
\everymath{\displaystyle}
\usepackage{mathtools}
\usepackage{amsmath,amsfonts,amssymb,amsthm}
\usepackage{graphicx}
\usepackage{float}
\usepackage{epstopdf}
\usepackage{booktabs}
\usepackage{tablefootnote}

\usepackage[authoryear,round]{natbib}
\bibpunct[, ]{(}{)}{,}{a}{}{,}

\usepackage{authblk}
\usepackage[hidelinks]{hyperref}

\begin{document}

\title{Dynamic slippage control and rejection feedback in spot FX market making}

\author{Alexander Barzykin\thanks{Email: \texttt{alexander.barzykin@hsbc.com}}}
\affil{HSBC, 8 Canada Square, Canary Wharf, London E14 5HQ, United Kingdom}
\date{\today}

\maketitle

\begin{abstract}
\noindent
We study an OTC FX market-making problem, built on the Avellaneda-Stoikov tradition, in which a dealer streams size-dependent quotes on a discrete ladder and manages inventory risk over a finite horizon under Poisson arrivals of trade requests.
Adverse selection is modeled through latency-driven price moves over a delay window, represented by Gaussian marks whose conditional means can depend on the quoted spread, capturing selective client reaction to stale quotes.
The dealer can address latency risk through trade rejection when slippage breaches a tolerance threshold.
We treat slippage tolerance as an explicit control jointly optimized with quotes: upon receiving a trade request, the dealer chooses an acceptance/rejection rule, which makes the trade economically akin to an embedded option written on the latency price move.
We further introduce rejection feedback through an EMA-based rejection score used as a reputation proxy, so that client intensity is endogenously modulated by past rejections via a multiplicative factor.
Using dynamic programming, we derive a Markov control problem with state variables (inventory, rejection-score) and show how rejection decision enters the Hamilton-Jacobi-Bellman equation through Hamiltonians that include an expectation over the latency mark and a maximization over both quote and rejection rule parameters.
For practical control evaluation, we develop an adiabatic-quadratic approximation: fixing reputation on the inventory-control time scale, expanding Hamiltonians to the second order {\`a} la Bergault {\itshape{et al.}}, and adopting quadratic ansatz in inventory, yielding tractable Riccati-type ODE and closed-form expressions for approximate quotes and slippage thresholds.
This approximation provides a fast surrogate for policy design and enables self-consistent calibration of rejection behavior.
We complement the analysis with exact numerical solutions of the reduced HJB and compare unrestricted and fair rejection protocols in terms of spread, rejection probability, and value.
\end{abstract}

\vspace{7mm}

\textbf{Keywords:} Foreign Exchange; Market Making; Stochastic Optimal Control; Market Impact; Algorithmic Trading; Price Validation; Trade Rejection; Last Look; Adverse Selection; Latency

\vspace{5mm}

\section{Introduction}

Electronic dealer-to-client (D2C) markets, especially in spot FX and related OTC products, routinely expose liquidity providers (LPs)
to latency risk: a client may hit a streamed quote that was formed on information that is already stale by the time the request reaches the dealer's pricing and risk stack.
A widely used response specific to FX is last look (LL), i.e., a short post-request window in which the dealer can accept or reject the trade, sometimes subject to protocol constraints on permissible price improvement or slippage.
As this practice reallocates execution risk between dealers and clients, its fairness relies on design choices such as
whether the acceptance rule is symmetric or asymmetric, whether price improvements are passed through, and how much discretion the dealer has.
These issues are central enough that industry codes emphasize transparency and the interpretation of LL as a risk-control mechanism rather than a profit tool 
\citep{gfxc2024}.

LL is primarily treated as an execution protocol shaping the client's effective costs and risks.
\citet{oomen2017b} develops a unified framework to compare LL designs -- symmetric versus asymmetric thresholds, and variants allowing or disallowing price
improvements/slippage -- focusing on execution risk, effective spreads, and how LL changes the trader's experience across protocols.
Closely related, \citet{stevenson2016} frames LL as a millisecond-expiry option embedded in the trading protocol and studies valuation and the incentives 
created by this opportunity.
\citet{oomen2017a} models execution through an aggregator that routes to the best displayed price among multiple LPs and highlights how competition for 
(apparently) uninformed flow can generate a ``winner's curse'' and adverse selection-type effects at the LP level.
\citet{cartea2019} develop an equilibrium-style model of FX trading with LL that distinguishes between fast and slow traders, and study how LL and competition
affect spreads, rejection rates, and welfare across participant types.
\citet{cartea2021a,cartea2021b} further study the interplay between latency and liquidity risk.

Current institutional practice reduces post-request hold time to zero -- the decision is made as soon as the request is received and resembles in this sense the mechanics of an exchange with firm liquidity.
The term ``last look'' itself may be misleading here, and one can talk about price validation instead.
There is an important difference from the exchange logic, however.
The dealer typically keeps slippage tolerance as a control parameter with an understanding that latency inherent in delocalized FX markets can lead to excessive rejections.

The market-making literature beginning with Avellaneda-Stoikov \citeyearpar{avellaneda2008} and subsequent refinements 
\citep{gueant2013,cartea2014,cartea2015,gueant2016} provides tractable stochastic-control foundations for optimal quoting under inventory risk 
and Poisson trade arrivals, but typically treats price dynamics as exogenous and does not model trade rejection decision as part of the control set.
Recently, \citet{olding2022} explicitly adapts classical market-making control models to incorporate LL via an exogenous survival/cancellation mechanism
(e.g., symmetric cancellation when the mid moves beyond a threshold during the LL window), deriving reservation-price adjustments and optimal quotes under LL.
Importantly, in Olding's control formulation LL parameters (window/threshold) are treated as given, while the dealer optimizes quotes.
Finally, beyond LL-specific work, the growing literature on informational risk in market making (e.g., alpha signals, quote-dependent adverse selection, price reading)
underscores that dealers' pricing choices can affect both the volume and the toxicity of the flow they attract \citep{cartea2020,barzykin2025}.

Despite this rich literature, two practical dimensions remain underdeveloped:
\begin{itemize}
\item {\bfseries Endogenous rejection policy design.} Dealers often choose rejection logic (e.g., a symmetric ``validity band'') as a dynamic risk-control policy that varies by market conditions, inventory, or relationship.
Much of the academic work either analyzes rejection as a static protocol feature (optionality/effective costs) or embeds slippage tolerance into market-making models with parameters fixed.
A natural step is to treat rejection decision -- subject to fairness/protocol constraints -- as an optimizable state-dependent control alongside quoting.

\item {\bfseries Intertemporal feedback from rejection decisions to future order flow.} Real OTC relationships are not memoryless: frequent rejections can reduce subsequent client engagement, routing priority, or hit rates.
Industry discussions emphasize transparency and appropriate use precisely because clients respond to perceived ``unfair'' or overly discretionary LL \citep{gfxc2024}.
Yet most tractable models do not include a reputational state variable that couples the current rejection decision to future flow intensity.

\end{itemize}

In this paper, we propose a tractable stochastic-control framework for D2C market making with latency that builds on the Avellaneda-Stoikov tradition, enhanced by optimal slippage tolerance with rejection feedback in response to latency risk.
We apply this framework to unconstrained rejection decisions and to one of the fair protocols, proposed by \citet{oomen2017b}, under strict zero hold-time policy (immediate decision upon request).
We also suggest an efficient adiabatic-quadratic approximation providing closed-form solutions, particularly in a simple but practically important case of 
constant adverse selection slippage.

\section{Unconstrained rejection as an arrival-time binary decision}

\subsection{Problem setup}

Let $S_t$ denote the reference mid price and assume that it undergoes the Brownian motion $W_t$ so that
\begin{equation}
dS_t = \sigma dW_t,
\end{equation}
with constant volatility $\sigma > 0$ on a finite operational horizon $[0, T]$ of the market maker (MM).
The MM streams OTC quotes to clients for a finite set of trade sizes $\{z_n\}_{n=1}^N \subset \mathbb{N}$ \citep{avellaneda2008,bergault2021b}.
For each bucket $n$ and side $i \in \{b, a\}$ (bid/ask), the MM controls the quote offsets $\delta_t^{n, i}$, so the executable quotes are
\begin{equation}
S_t^{n,i} = S_t \mp \delta_t^{n,i}, \qquad i \in \{b, a\}.
\end{equation}
Client requests arrive according to counting processes $M^{n,i}_t$ with intensities
\footnote{Exponential intensity is a natural assumption, but sigmoid is also used, particularly for RFQ \citep{barzykin2023}.}
\begin{equation}
\lambda^{n,i}_t = g(R_t) \Lambda^n_0\,e^{-\kappa_n\delta^{n,i}_t},
\end{equation}
where $\Lambda^n_0 > 0$, $\kappa_n > 0$ and $g : [0, 1] \to \mathbb{R}^+$ is a decreasing function of a rejection-score variable $R_t$ (a reputation proxy) constructed as an
exponentially weighted moving average (EMA) of past rejects.\footnote{One can easily verify whether the effect is present in the system by regressing the number of requests over a period of time (say, one hour) versus the average realized rejection rate over the previous period.}

A request of type $(n, i)$ arriving at time $t$ is a response to a quote $\delta$ that is $\tau_n$ old.
Let $Y^{n,i}_t(\delta)$ denote the sign-adjusted mid-price change over that latency window (positive move is in the dealer's favour).
We assume a Gaussian mark with a quote-dependent mean:
\begin{equation}
Y^{n,i}_t(\delta) = m_n(\delta) + \nu_n \xi, \qquad \xi \sim \mathcal{N}(0, 1), \qquad \nu_n := \sigma \sqrt{\tau_n} .
\end{equation}
where $m_n(\delta)$ is a measurable function encoding adverse selection.\footnote{Exponential saturation is a natural default \citep{barzykin2025}.}
In what follows, we assume side-symmetric latency marks: the conditional law of $Y^{n,i}_t$ depends on the side only through the sign convention used to define favourable and adverse moves, while the functions $m_n(\cdot)$ and $\nu_n$ are common to bid and ask.
The mark is observed by the MM prior to accepting or rejecting the trade.
Hereinafter, we assume zero hold time by the dealer, i.e., the decision $\ell^{n,i}_t \in \{0, 1\}$ is made as soon as the trade request is received.
Here $\ell = 1$ denotes acceptance and $\ell = 0$ rejection.
Reputation (rejection score) is updated per request
\begin{equation}
R^+ = (1 - \rho) R + \rho (1 - \ell), \qquad \rho \in (0, 1) ,
\end{equation}
with maps
\begin{equation}
R_\text{acc}(R) = (1-\rho)R, \qquad R_\text{rej}(R) = (1-\rho)R + \rho .
\end{equation}
Thus $R_t \in [0, 1]$ whenever $R_0 \in [0, 1]$.
\footnote{An alternative way to define the reputation variable is via an Ornstein-Uhlenbeck process driven by rejections.}

Let $q_t \in \mathbb{Z}$ denote inventory and $X_t \in \mathbb{R}$ cash.
When a bid request is accepted the MM buys $z_n$ at $S_t - \delta^{n,b}_t$; when an ask request is accepted the MM sells $z_n$ at $S_t + \delta^{n,a}_t$.
Hence
\begin{equation}
dq_t = \sum_{n=1}^N z_n \ell^{n,b}_t\, dM^{n,b}_t - \sum_{n=1}^N z_n \ell^{n,a}_t\, dM^{n,a}_t ,
\end{equation}
\begin{equation}
dX_t = -\sum_{n=1}^N \ell^{n,b}_t z_n (S_t - \delta^{n,b}_t)\, dM^{n,b}_t + \sum_{n=1}^N \ell^{n,a}_t z_n (S_t + \delta^{n,a}_t)\, dM^{n,a}_t .
\end{equation}
The dealer aims to maximize expected terminal mark-to-market wealth penalized by running inventory risk \citep{cartea2014}:
\begin{equation}
\sup_{\delta, \ell} \mathbb{E} \left[ X_T + q_T S_T - \frac{1}{2} \gamma \sigma^2 \int_0^T q_t^2 dt \right], \qquad \gamma > 0 ,
\end{equation}
possibly with a terminal penalty.

\subsection{HJB and optimal controls}

Define the value function on the full state $U(t, x, q, R, S)$.
The corresponding Hamilton-Jacobi-Bellman (HJB) equation reads:
\begin{equation}
0 = \partial_t U + \frac{1}{2} \sigma^2 \partial_{SS}U - \frac{1}{2}\gamma \sigma^2 q^2
+ \sum_{i \in \{b,a\}} \sum_{n=1}^N \sup_{\delta^{n,i}} \lambda^{n,i} \mathbb{E} \left[ \sup_\ell \big(U^{n,i}_0(\ell) - U\big)\right] ,
\end{equation}
where $U^{n,i}_0(\ell)$ denotes the value immediately after processing the request of type $(n,i)$ under decision $\ell$ and expectation
is over the Gaussian mark $Y^{n,i}(\delta)$.

Since the objective is linear in cash and mark-to-market wealth, we employ the affine ansatz
\begin{equation}
U(t, x, q, R, S) = x + qS + V(t, q, R)
\end{equation}
for the reduced value function $V$.
This removes the mid-price state from the HJB.
For a request of size bucket $n$, acceptance generates incremental value
\begin{equation}
\Delta_\text{acc}^{n,i} = z_n \left( \delta + Y^{n,i}(\delta^{n,i}) - \mathcal{D}^n_\pm V(t, q, R) \right)
\end{equation}
while rejection generates
\begin{equation}
\Delta_\text{rej}^{n,i} = \mathcal{J}_R V(t, q, R) ,
\end{equation}
where
\begin{equation}
p^{n,i}(t, q, R) \equiv \mathcal{D}^n_{\pm} V(t, q, R) := \frac{V\big(t, q, R\big) - V\big(t, q \pm z_n, R_\text{acc}(R)\big)}{z_n} ,
\end{equation}
\begin{equation}
J(t, q, R) \equiv \mathcal{J}_R V(t, q, R) := V(t,q,R_\text{rej}(R)) - V(t,q,R) .
\end{equation}
Thus, the optimal decision induces a pointwise maximum.
Hence, the reduced HJB for $V(t,q,R)$ can be written as
\begin{equation}
0 = \partial_t V - \frac{1}{2}\gamma \sigma^2 q^2 
+ g(R) \sum_{i \in \{b,a\}} \sum_{n=1}^N \sup_{\delta^{n,i}} \Lambda^n_0\,e^{-\kappa_n\delta^{n,i}} 
\mathbb{E} \left[ \max \Big( \Delta_\text{acc}^{n,i}, \Delta_\text{rej}^{n,i} \Big)\right] ,
\end{equation}
with terminal condition $V(T, q, R) = \mathcal{T}(q, R)$.\footnote{We are primarily interested in the near-stationary solution, so the terminal condition can be set to zero.}
Note that since $g(R)$ is decreasing and a rejection moves the score from $R$ to $R_\text{rej}(R) \ge R$, it is natural to expect the value function to be non-increasing in $R$.
Under this monotonicity, $J(t, q, R) \le 0$.
We can interpret $J$ as the continuation loss from worsening reputation.

\begin{figure}[ht]
\centering
\includegraphics[width=0.7\columnwidth]{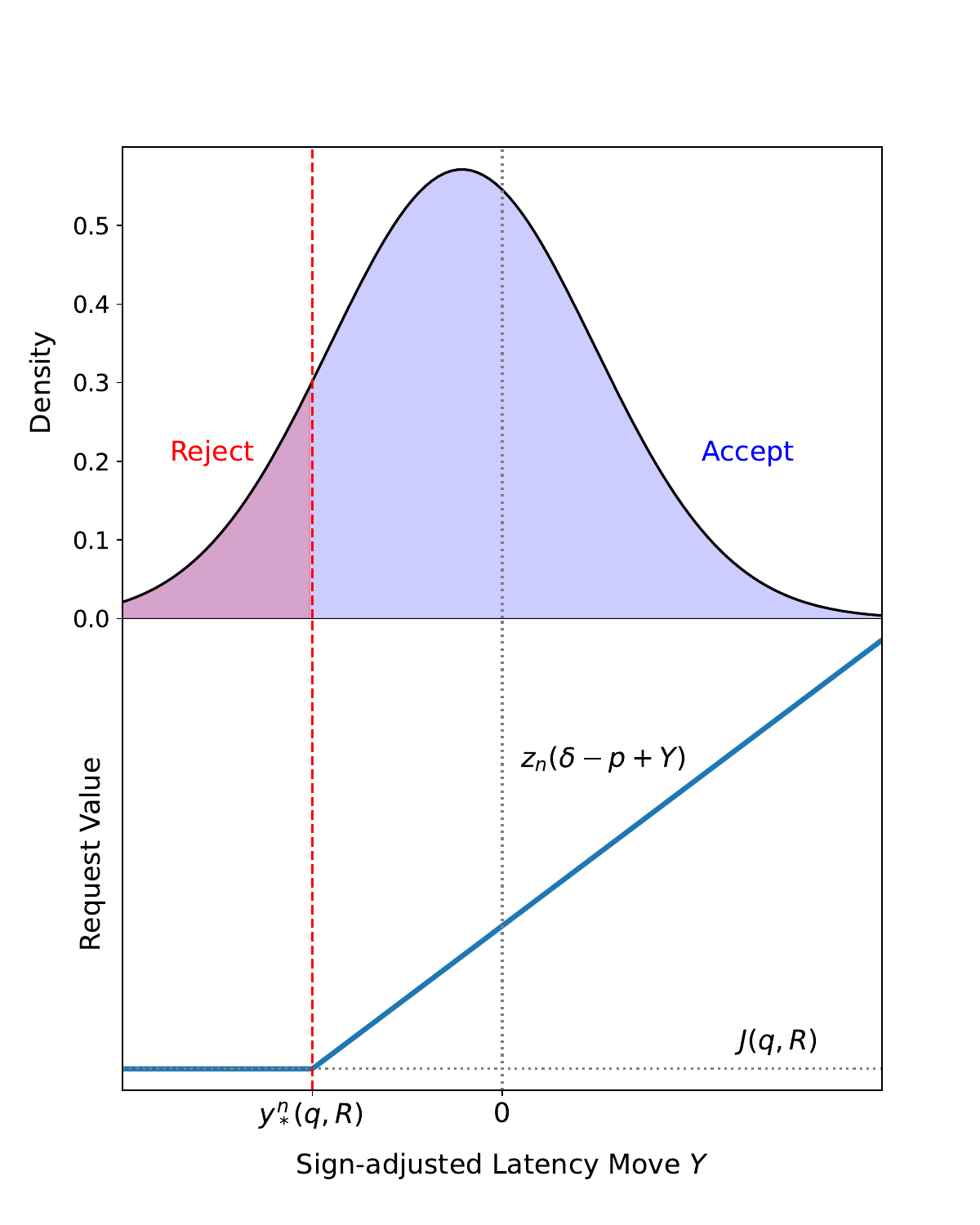}
\caption{
Latency mark $Y \sim \mathcal{N}(m_n(\delta), \nu_n^2)$ and embedded-option view of request-level payoff for unrestricted rejection policy.
Red dashed line depicts optimal decision threshold: the request is accepted when $Y >= y_*^n(q, R)$.
}
\label{call_option}
\end{figure}

Let $\Phi$ and $\phi$ denote the standard normal CDF and PDF, respectively, and define
\begin{equation}
\Psi_n(\mu) := \mathbb{E}\left[\big(\mathcal{N}(\mu+\nu_n \xi)\big)_+\right] = \nu_n \phi(\mu/\nu_n) + \mu \Phi(\mu/\nu_n) .
\end{equation}
Using $\max(A, B) = B + (A-B)_+$, we obtain
\begin{equation}
\mathbb{E} \left[ \max \Big( z_n(\delta + Y - p), J \Big)\right] = J + z_n \Psi_n\left(\delta + m_n - p - \frac{J}{z_n}\right)
\end{equation}
when $Y \sim N(m_n, \nu_n^2)$.
Here the second term can be identified as the value of exercising the accept option, as illustrated in Figure \ref{call_option}.
If we define for each size bucket $n$ the bid/ask Hamiltonians
\begin{equation}
H^{n}(p, J) := \sup_{\delta \in \mathbb{R}} \Lambda_0^n e^{-\kappa_n \delta} 
\left[J + z_n \Psi_n \left(\delta + m_n(\delta) - p - \frac{J}{z_n}\right)\right] ,
\end{equation}
we can further transform the reduced HJB to
\begin{equation}
\label{reducedHJB}
0 = \partial_t V - \frac{1}{2}\gamma \sigma^2 q^2 + g(R) \sum_{i \in \{b,a\}} \sum_{n=1}^N
H^{n}\left( p^{n,i}(t, q, R), J(t, q, R)\right) .
\end{equation}

For each $(n, i)$, optimal quotes solve the $\sup_\delta$ in HJB:
\begin{equation}
\delta^{n,i}_*(t, q, R) = \text{arg} \max_\delta \Lambda^n_0 e^{-\kappa_n \delta} 
\left[J(t, q, R) + z_n \Psi_n \left(\delta + m_n(\delta) - p^{n,i}(t, q, R) - \frac{J(t, q, R)}{z_n}\right) \right] .
\end{equation}
For an interior maximizer $\delta^* = \delta^*_n(p, J)$, the first-order condition (FOC) is
\begin{equation}
z_n \Phi\left(\frac{\mu_n(\delta^*;p,J)}{\nu_n}\right)\left(1 + m'_n(\delta^*)\right) = \kappa_n
\left(J + z_n \Psi_n\big(\mu_n(\delta^*;p,J)\big)\right) ,
\end{equation}
where
\begin{equation}
\mu_n(\delta;p,J) = \delta + m_n(\delta) - p - \frac{J}{z_n} .
\end{equation}
When $J \equiv 0$ and $m^n \equiv 0$ and $\nu_n \to 0$, the model collapses to the standard OTC HJB (without rejection control) and the familiar
$\delta^{n,i}_* = 1/\kappa_n + \mathcal{D}^n_\pm V$ regime \citep{gueant2013}.

The optimal decision arises from the comparison of the accept vs. reject value increments:
\begin{equation}
\ell^{n, i}_* = 1 \quad \text{iff} \quad
z_n \left( \delta^{n,i}_* + Y^{n,i}(\delta^{n,i}_*) - p^{n,i}\right) \ge J(t, q, R) .
\end{equation}

\subsection{Adiabatic quadratic approximation}

In our practical definition, the rejection score $R$ is an EMA of the rejection indicator and, therefore, tracks the rejection probability.
It is intentionally filtered and is expected to change slowly, maintaining stationarity.
The full problem thus couples a fast inventory control with a slow reputation state.
In the adiabatic regime, we approximate $R$ as quasi-static over the quoting horizon, set $g(R) =: g$ and drop $R$-dependence from the value function.
In the self-consistent closure, we replace the continuation difference by a constant $J \le 0$ calibrated to match the corresponding rejection rate. 
HJB becomes inventory-only.

Following \citet{bergault2021a} we expand the Hamiltonians in $p$ up to the second order
\begin{equation}
H^n(p,J) \simeq H^n(0, J) + H^n_p(0, J) p + \frac{1}{2} H^n_{pp}(0,J) p^2 ,
\end{equation}
and assume a standard symmetric quadratic ansatz for the value function
\begin{equation}
V(t,q) = -A(t) q^2 - C(t) .
\end{equation}
Substituting into HJB and matching the coefficients of $q^2$ yields the Riccati ODE for $A(t)$:
\begin{equation}
A'(t) + \frac{1}{2} \gamma \sigma^2 = 4 g A(t)^2 \Sigma(J) , \qquad \Sigma(J) := \sum_{n=1}^N H^n_{pp}(0, J) .
\end{equation}
In the stationary (long-horizon) limit $A(t)' = 0$, the concave branch gives
\begin{equation}
A = \sqrt{\frac{\gamma \sigma^2}{8g\Sigma(J)}} .
\end{equation}
A free constant does not influence the controls.
Given $A$, one can compute $p^{n,i}(q) = A(z_n \pm 2q)$ and obtain quotes by solving FOC.

\subsubsection*{Special case: constant slippage}

We now specialize the adiabatic-quadratic approximation to constant (quote-independent) slippage:
\begin{equation}
m_n(\delta) \equiv -\theta_n, \qquad \theta_n \ge 0 ,
\end{equation}
which yields explicit forms for Hamiltonian derivatives and particularly simple controls.
In this case,
\begin{equation}
H^n(p,J) = \Lambda_0^n e^{-\kappa_n \delta} 
\left[J + z_n \Psi_n \left(\delta -\theta_n - p - \frac{J}{z_n}\right)\right] .
\end{equation}
Setting $u = \delta - p$ yields
\begin{equation}
H^n(p, J) = e^{-\kappa_n p} H^n(0, J) ,
\end{equation}
and, therefore, the maximizer satisfies the exact shift rule
\begin{equation}
\delta^*_n(p, J) = \bar{\delta}_n(J) + p , \qquad \bar{\delta}_n(J) := \delta^*_n(0, J) .
\end{equation}
Define
\begin{equation}
\bar{\mu}_n(J) := \bar{\delta}_n(J) - \theta_n - \frac{J}{z_n} , \qquad \tilde{\mu}_n(J) := \frac{\bar{\mu}_n(J)}{\nu_n} .
\end{equation}
The myopic optimizer at $p = 0$ satisfies
\begin{equation}
\Phi(\tilde{\mu}_n) = \kappa_n \left( \frac{J}{z_n} + \nu_n \big(\phi(\tilde{\mu}_n) + \tilde{\mu}_n \Phi(\tilde{\mu}_n)\big)\right) ,
\end{equation}
and once solved for $\tilde{\mu}_n$,
\begin{equation}
\bar{\delta}_n(J) = \theta_n + \frac{J}{z_n} + \nu_n \tilde{\mu}_n(J) .
\end{equation}
Using the FOC at the optimizer, we find
\begin{equation}
H^n_{pp}(0, J) = \Lambda^n_0 e^{-\kappa_n \bar{\delta}_n(J)} z_n \kappa_n \Phi\big(\tilde{\mu}_n(J)\big) .
\end{equation}
Hence
\begin{equation}
\Sigma(J) = \sum_{n=1}^N \Lambda_0^n e^{-\kappa_n \bar{\delta}_n(J)} z_n \kappa_n \Phi\big(\tilde{\mu}_n(J) \big) .
\end{equation}
By the shift rule, the adiabatic optimal quotes become explicit
\begin{equation}
\label{oquotes}
\delta^{n,i}_* = \bar{\delta}_n(J) + A (z_n \pm 2 q) ,
\end{equation}
and for trade acceptance/rejection controls we have
\begin{equation}
\ell^{n,i}_* = 1 \quad \text{iff} \quad Y^{n,i} \ge y^{n,i}_* := \frac{J}{z_n} - \bar{\delta}_n(J) .
\end{equation}

\subsubsection*{Calibration of $J$ by average rejection rate}

At $p = 0$, the reject probability under the optimal quote in size bucket $n$ is $1 - \Phi\big(\tilde{\mu}_n(J)\big)$.
A natural intensity-weighted average rejection rate is, therefore,
\begin{equation}
\bar{r}(J) = \frac{\sum_{n=1}^N \Lambda_0^n  e^{-\kappa_n \bar{\delta}_n(J)} \left(1 - \Phi\big(\tilde{\mu}_n(J)\big)\right)}
{\sum_{n=1}^N \Lambda_0^n  e^{-\kappa_n \bar{\delta}_n(J)}} .
\end{equation}
Then we set $J$ such that $\bar{r}(J) = R$.
Interpretationally, $J$ plays the role of a shadow continuation loss associated with a rejection.
Under slow reputation dynamics, we solve the fast inventory-control problem at frozen $R$, compute the resulting stationary reject frequency $\bar{r}(J)$,
and choose $J$ so that this stationary reject frequency matches the current reputation level.
This provides a self-consistent local closure linking the fast control problem to the slow reputation state.

\begin{figure}[ht]
\centering
\includegraphics[width=0.7\columnwidth]{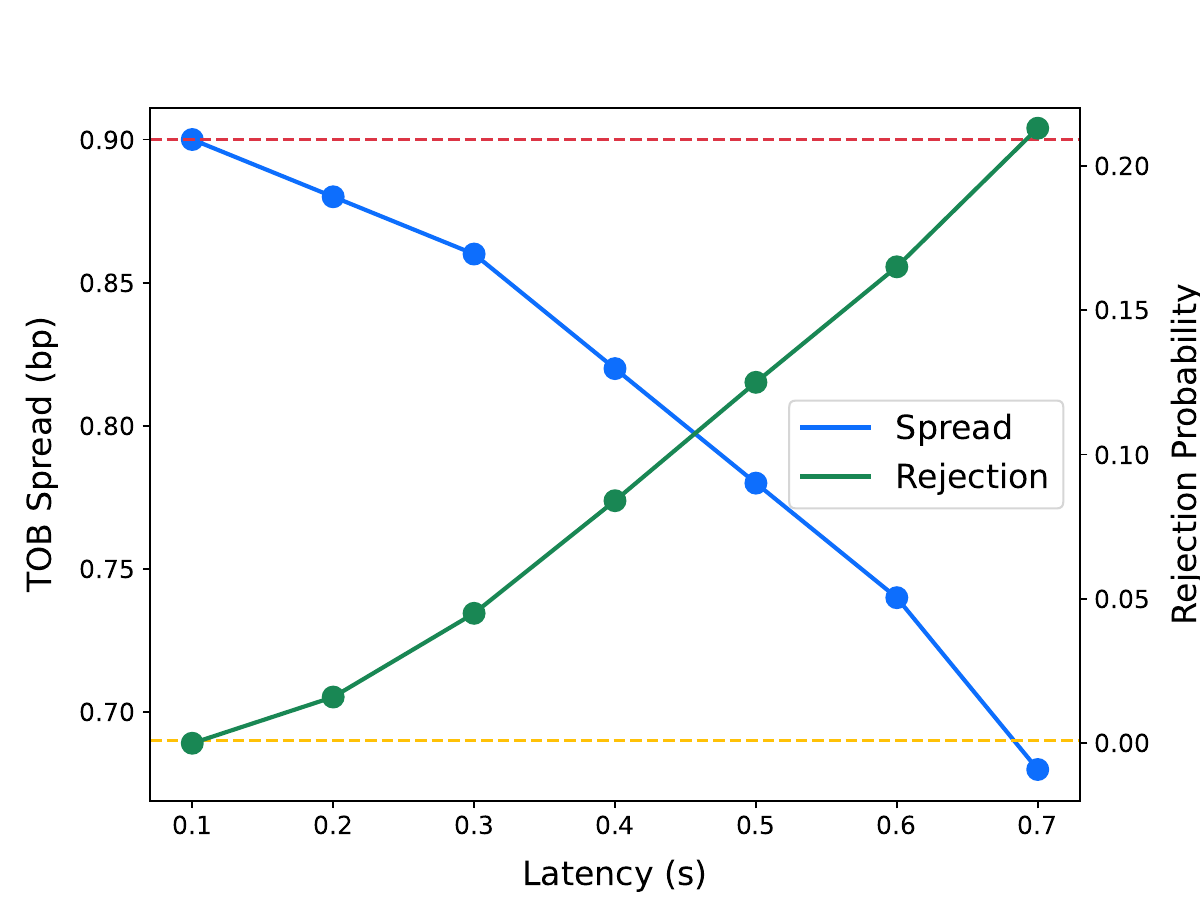}
\caption{
Top of book ($z = z_1$) spread and rejection probability of optimal MM facing toxic flow with unrestricted rejection and no reputation feedback as a function of latency.
The top dashed line corresponds to the spread with toxic flow, but no rejection, and the bottom dashed line shows the spread without toxicity and without rejection.
Parameters in the text.
}
\label{latency1d}
\end{figure}

\begin{figure}[ht]
\centering
\includegraphics[width=0.7\columnwidth]{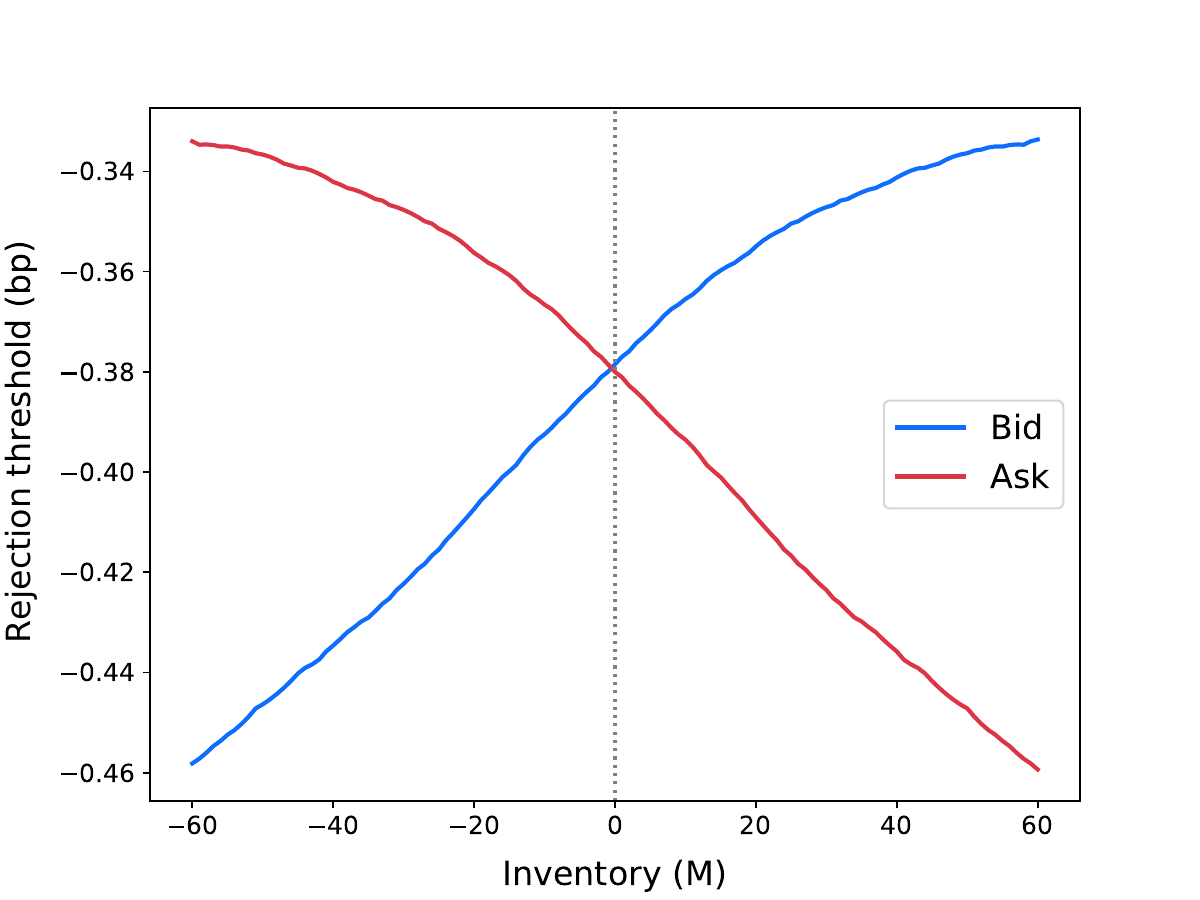}
\caption{
Bid and ask unrestricted rejection thresholds for trade requests of size $z_1$ as functions of inventory for optimal MM facing flow with toxicity function 
$m_n(\delta) = -0.1 \exp(-0.5\delta)$ and latency window $\tau_n = 0.5$ s.
Other parameters in the text.
}
\label{threshold1d}
\end{figure}

\subsection{Numerical examples}

As an illustration, consider a standard size-ladder of $\Delta^n = (1, 2, 5, 10, 20)$ M notional and an exponential intensity function 
with $\Lambda_0^{n} = (2000, 800, 600, 400, 100)$~day$^{-1}$ and $\kappa_n = (3.0, 2.5, 2.0, 1.5, 1.0)$ bp$^{-1}$.
Here bp stands for basis points.\footnote{This implies GBM while we deal here with simple Brownian motion. 
The difference is negligible in FX market making due to short trading horizons.}
This set of parameters corresponds to a liquid currency pair with a daily turnover of $ \simeq 4$ billion notional and a top-of-book spread of $\simeq 0.7$~bp. 
We also assume a daily volatility of $100$~bp and a risk aversion coefficient of $\gamma = 10^{-3}$~bp$^{-1}$~M$^{-1}$.

We begin without rejection feedback and set a small constant toxicity of $\theta_n = 0.1$ bp across sizes.
Figure~\ref{latency1d} demonstrates that if the quote-to-received-request latency is capped at 100 ms, there is no point for the dealer to reject at these parameters.
\footnote{The corresponding one-dimensional HJB is solved numerically.}
Understandably, the optimal spread is wider than without toxicity.
With increasing latency, the MM starts rejecting, and the spread tightens, a well-known phenomenon \citep{oomen2017a, oomen2017b, cartea2019}.
Rejection thresholds are not sensitive to inventory in this constant-toxicity setting.
To introduce inventory dependence, we next use a quote-dependent toxicity specification $m_n(\delta) = -\theta_n e^{-\beta_n \delta}$,
and set $\beta_n = 0.5$ bp$^{-1}$, as shown in Figure~\ref{threshold1d}.
Rejection thresholds remain relatively deep, indicating that even an unconstrained MM does not reject immediately upon observing adverse slippage.

\begin{figure}[ht]
\centering
\includegraphics[width=0.7\columnwidth]{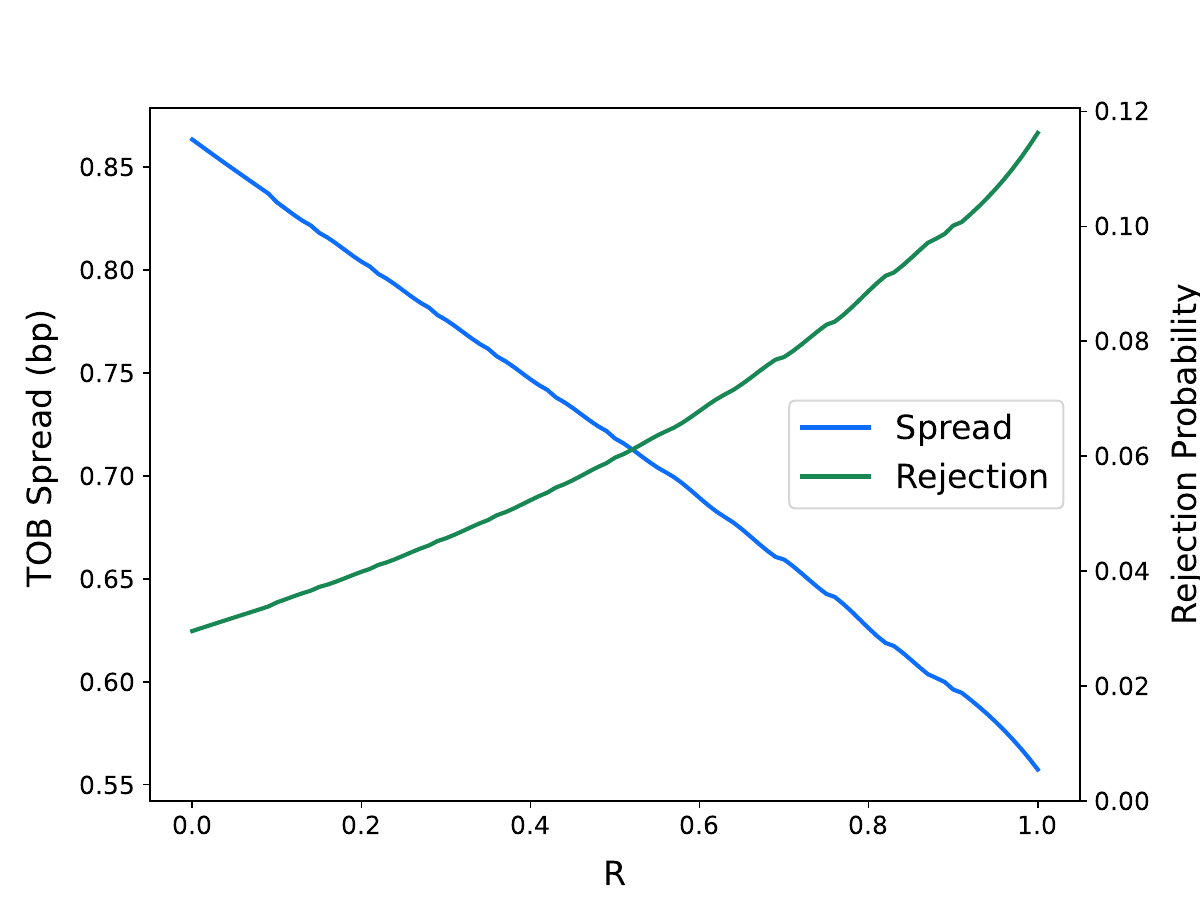}
\caption{
Optimal top of book spread and rejection probability as functions of the rejection score $R$ for zero inventory.
Parameters in the text.
}
\label{optimal1d}
\end{figure}
\begin{figure}[ht]
\centering
\includegraphics[width=0.7\columnwidth]{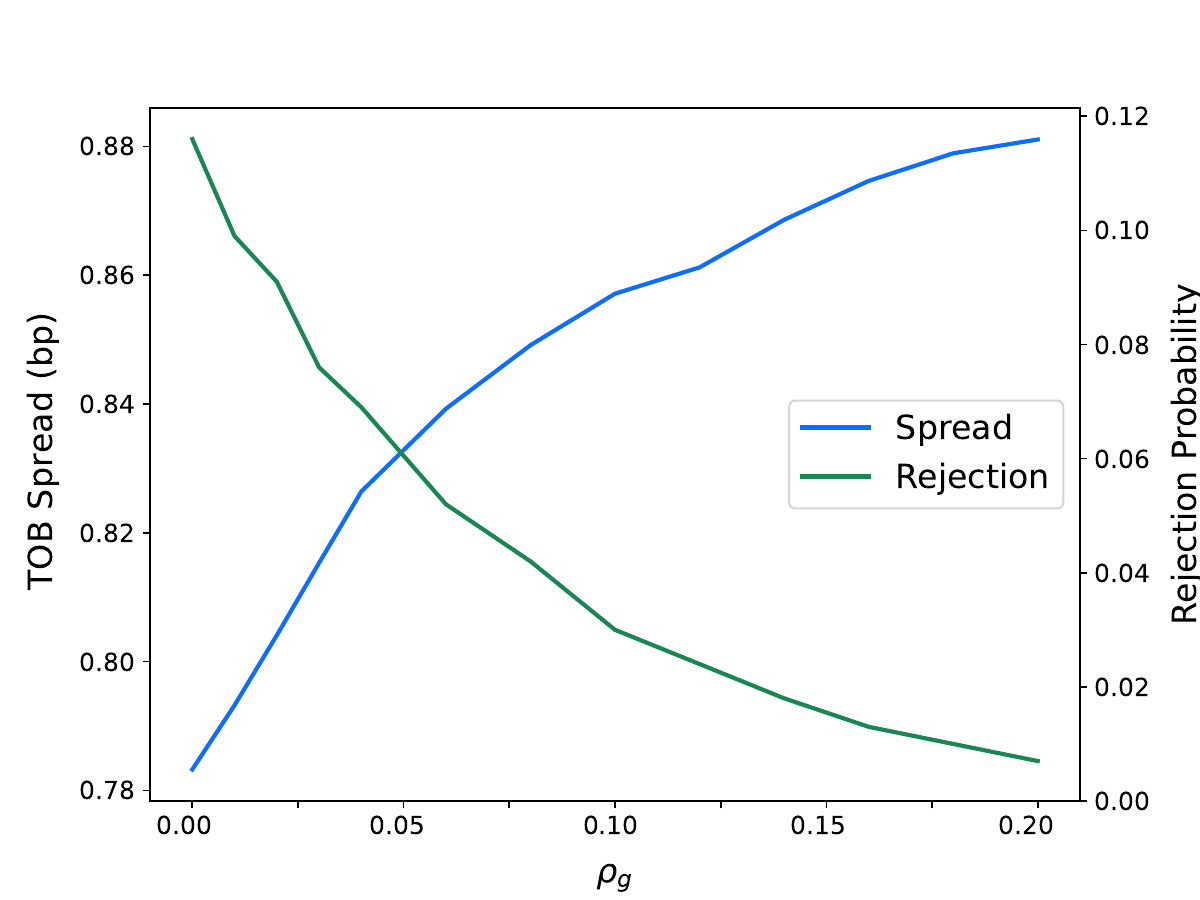}
\caption{
Stationary top of book spread and rejection probability as functions of the rejection score decay parameter $\rho_g$, such that $g(R) = \exp(-\rho_g R)$.
Parameters in the text.
}
\label{reputation_decay1d}
\end{figure}

Now consider optimal market making with rejection feedback.
We set EMA smoothing parameter to $\rho = 0.01$ and let $g(R) = \exp(-\rho_g R)$ with $\rho_g = 0.1$.
We also assume that $\tau_n = 0.5$ s and $\beta_n = 0.3$ bp$^{-1}$.
All other parameters are the same as defined above.
Figure~\ref{optimal1d} illustrates a mechanism that may appear counter-intuitive at first: when the rejection score $R$ is high, the dealer tightens spreads yet becomes more selective ex post.
The reason is that, under adiabatic approximation, a higher $R$ reduces future flow through $g(R)$, while the continuation loss from an additional rejection becomes smaller in magnitude.
Tighter quoting then helps restore reputation faster through a higher arrival rate and hence a faster dilution of past rejections in calendar time.
This ``repair-exploit'' incentive to tighten in order to restore reputation faster while using the optionality of unconditional rejection to filter losses is a
structural feature of the model and provides a motivation for considering fair protocols that attenuate the embedded one-sided option value.
\footnote{Changing $R$ to be driven by rejections alone can help as well.}

At the same time, fast reputation repair implies staying close to stationarity.
From the EMA update, the conditional mean change per request is $\rho (\bar{r} - R)$, where $\bar{r}$ is the intensity-weighted reject probability.
Thus in calendar time $\dot{R} \approx \rho \lambda(q, R) \left( \bar{r}(q, R) - R\right)$ with $\lambda(q, R) = \sum \lambda^{n,i}(q, R)$.
Stationary reputation levels satisfy the fixed-point relation (for fixed inventory $q$)
\begin{equation}
R^* = \bar{r}(q, R^*) ,
\end{equation}
which is independent of $\lambda$, while $\lambda \rho$ controls the speed of convergence to $R^*$.
Figure~\ref{reputation_decay1d} demonstrates that in the stationary state, the rejection probability actually decreases and the spread widens 
when we increase the reputation-induced flow decay parameter $\rho_g$, as expected.
In other words, the dealer will try to avoid rejection on average, even unconditionally, in the presence of reputation feedback.

\begin{figure}[ht]
\centering
\includegraphics[width=0.7\columnwidth]{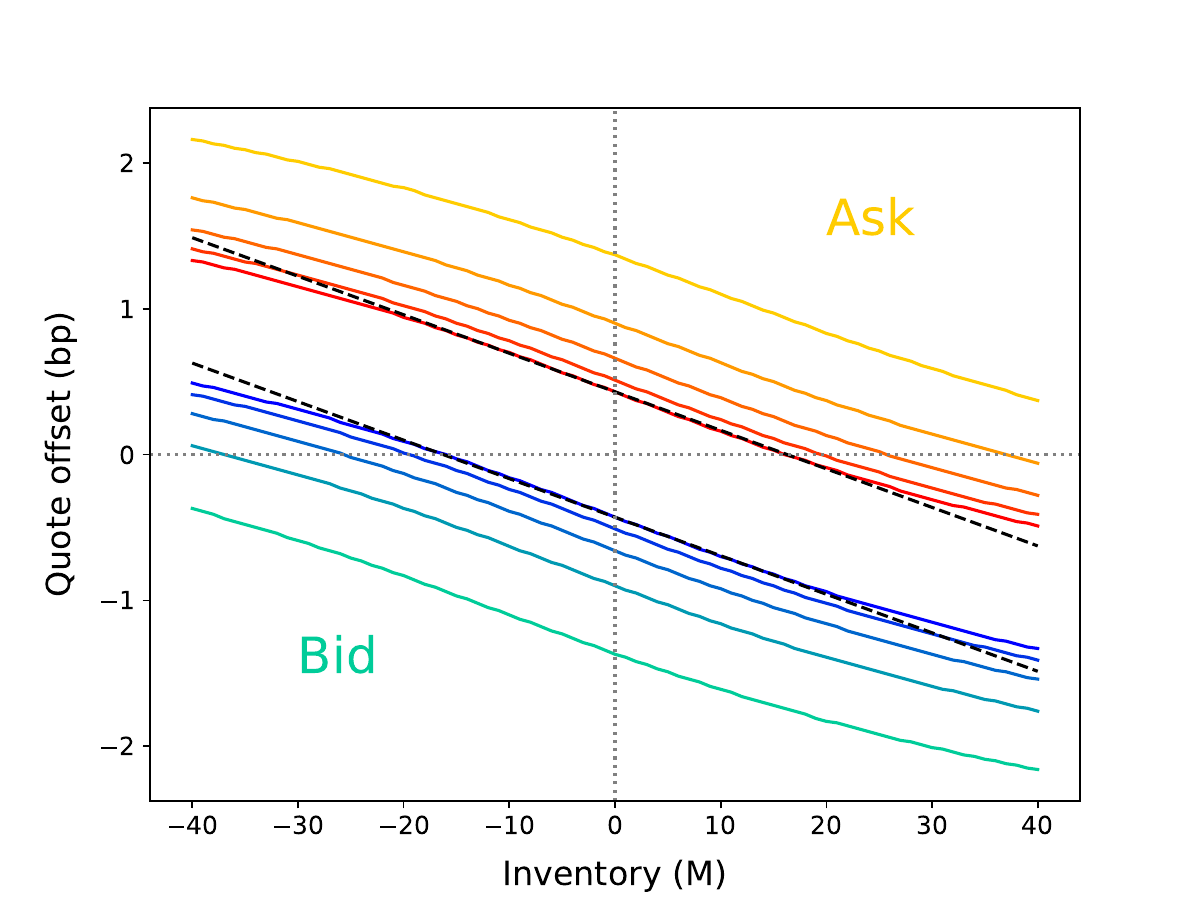}
\caption{
Optimal pricing as a function of inventory under a stationary rejection score.
Parameters in the text.
Lighter colour shades correspond to larger values of ${z_n}$ in the ladder.
Dashed lines show adiabatic quadratic approximation for size $z_1$.
}
\label{quotes1d}
\end{figure}

Finally, we check the validity of the adiabatic quadratic approximation.
For our parameter set (with $\rho_g = 0.1$), the equilibrium reputation level is $R^* \simeq 0.03$.
This is the value we use to evaluate pricing in Figure~\ref{quotes1d}.
The accuracy is reasonable, particularly for small inventories (comparison is only demonstrated for the smallest size bucket $z_1$).
We have also verified rejection thresholds and found a good correspondence (not shown).

\section{Towards fair execution}

Reputation feedback ensures that excessive rejection practice is penalized.
One can even introduce dedicated reputation-dependent penalties into the utility function to gain more control.
However, the question of rejection fairness will still remain because the decision is with the dealer while the opportunity cost is with the client,
particularly if the dealer rejects adverse selection but harvests favorable selection.
A number of protocols have been discussed in detail by \citet{oomen2017b} offering fairness via decision symmetry and/or fill price adjustments.
Under certain scenarios, the benefit of rejection can totally disappear, at least theoretically,\footnote{
For example, under quadratic inventory penalty and symmetric rejection boundary without adverse selection, as demonstrated by \citet{olding2022}.}
so the modelling framework is as important as the policy design.

So far, we have defined slippage tolerance as a function of the reference mid-price move over the latency window.
The approach is aligned with ``fairness'' narratives, yet in most OTC FX implementations, the price check is expressed
against the dealer's own price stream.
This is precisely the object used in standard protocol definitions \citep{oomen2017a}.
The goal of this section is to replace mid-move marks by dealer-quote-move marks and embed fair execution protocols into the
dynamic control framework with reputation feedback and inventory risk.

We begin by introducing the sign-adjusted quote move
\begin{equation}
Y^{n,i}_t := \pm \left( S^{n,i}_t - S^{n,i}_{t-\tau_n}\right) ,
\end{equation}
so that $Y > 0$ means the move is in the dealer's favour, as before.
We retain the Gaussian-mark assumption (now interpreted as a distribution for own-quote revisions):
\begin{equation}
Y^{n,i}(\delta) = m_n(\delta) + \nu_n \xi, \qquad \xi \sim \mathcal{N}(0, 1), \qquad \nu_n = \sigma \sqrt{\tau_n} .
\end{equation}
At this point, one should note that the quote move may in principle contain both exogenous market motion and endogenous quote revisions generated by the dealer's own policy updates.
In the present tractable formulation, we treat the valudation mark $Y^{n,i}_t$ as an exogenous effective quote-revision variable with conditional law $m_n(\delta) + \nu_n \xi$.
Equivalently, one may view $S^{n,i}$ here as the reference quote stream used for validation,\footnote{Desks often maintain reference streams.} with the control frozen over the latency window on the decision scale.
This approximation preserves the dynamic-programming structure while aligning the validation rule with standard OTC protocol definitions.
All the reputation/inventory machinery from the previous section remains unchanged; only the economic mapping from $Y$ to execution P\&L
is modified to reflect the chosen protocol.

\subsection{Symmetric tolerance with capped slippage}

\begin{figure}[h!]
\centering
\includegraphics[width=0.7\columnwidth]{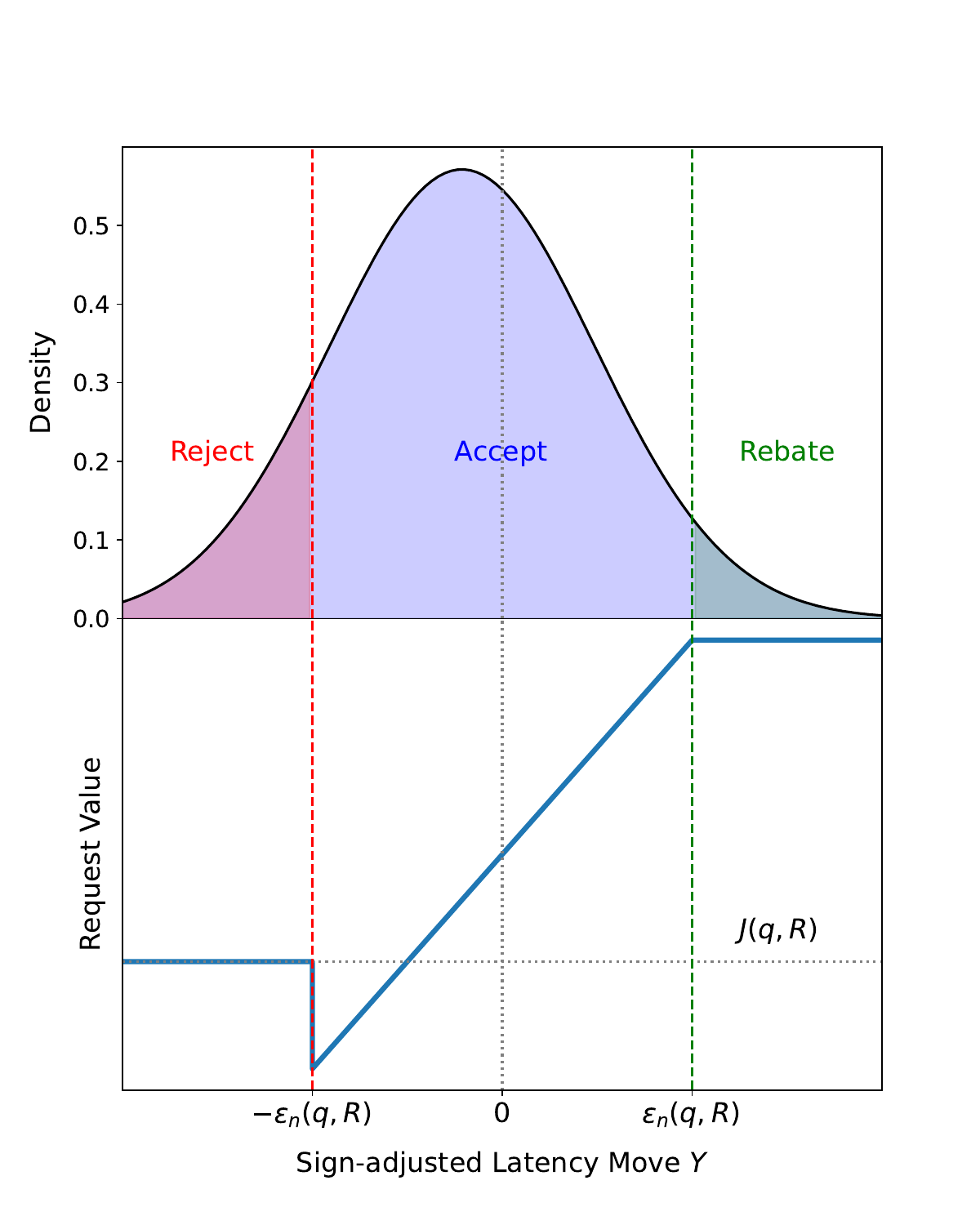}
\caption{
Latency mark $Y \sim \mathcal{N}(m_n(\delta), \nu_n^2)$ and embedded-option view of request-level payoff for fair rejection policy as described in the text.
For clarity, side indices are suppressed in the figure; in the model, the corresponding marginal value $p^{n,i}(t, q, R)$ remains side dependent.
Red dashed lines depict optimal decision thresholds: the request is rejected when $Y < -\epsilon^n(q, R)$; when $Y > \epsilon^n(q, R)$,
the request is accepted with price improvement for the client.
}
\label{barrier_option}
\end{figure}

Following \citet{oomen2017b}, we consider the execution policy with symmetric tolerance $\epsilon$:
\begin{itemize}
\item Reject if $Y < -\epsilon$ .
\item Accept on-rate if $-\epsilon \le Y \le \epsilon$ (dealer keeps the realized slippage).
\item Accept with rebate if $Y > \epsilon$ (dealer's favourable slippage is capped at $\epsilon$).
\end{itemize}
Equivalently, conditional on acceptance, the dealer's slippage term becomes $\min(Y, \epsilon)$ with
$\epsilon$ being one of the dealer's controls.
This protocol is not statistically symmetric in the realized slippage, but it is economically symmetric in the sense that extreme favorable slippage is rebated while extreme adverse slippage leads to rejection.
It is therefore comparatively easy to communicate and defend, while remaining rich enough to illustrate the optimal-control and embedded-option viewpoints, as shown in Figure~\ref{barrier_option}.

At state $(t, q, R)$ we define, as before, $J = J(t, q, R) = V(t,q,R_\text{rej}(R)) - V(t,q,R)$ and $p = p^{n,i}(t, q, R)$
being the appropriate accept marginals $\mathcal{D}^n_{\pm} V(t, q, R)$.
Given $(\delta, \epsilon)$, a single request contribution (inside the intensity-weighted Hamiltonian) is
\begin{equation}
J \cdot \mathbf{1}_{Y < - \epsilon} + z_n \left(\delta - p + \min(Y, \epsilon)\right) \cdot \mathbf{1}_{Y \ge-\epsilon} .
\end{equation}
Define the standardized cutoffs
\begin{equation}
a_n = \frac{- \epsilon -  m_n}{\nu_n} , \qquad b_n = \frac{\epsilon -  m_n}{\nu_n}
\end{equation}
Then
\begin{equation}
\mathbb{P}(Y < -\epsilon) = \Phi(a_n), \qquad \mathbb{P}(-\epsilon \le Y \le \epsilon) = \Phi(b_n) - \Phi(a_n),
\qquad \mathbb{P}(Y > \epsilon) = 1 - \Phi(b_n) ,
\end{equation}
and
\begin{equation}
\mathbb{E}\left[Y \cdot \mathbf{1}_{-\epsilon \le Y \le \epsilon} \right] =  m_n \left(\Phi(b_n) - \Phi(a_n)\right)
+ \nu_n \left( \phi(a_n) - \phi(b_n) \right) .
\end{equation}
So the closed-form expected increment is
\begin{equation}
\mathcal{G}_n(\delta,\epsilon;p, J) = J \Phi(a_n) + z_n (\delta - p) B_n + z_n C_n ,
\end{equation}
where we have introduced
\begin{equation}
B_n = 1 - \Phi(a_n), \qquad C_n = m_n \left(\Phi(b_n) - \Phi(a_n)\right) + \nu_n \left( \phi(a_n) - \phi(b_n) \right) + \epsilon \left(1 - \Phi(b_n)\right) .
\end{equation}
The bucket Hamiltonian becomes
\begin{equation}
H^{n}(p, J) := \sup_{\delta \in \mathbb{R}, \epsilon \ge 0} \Lambda_0^n e^{-\kappa_n \delta} 
\mathcal{G}_n(\delta,\epsilon;p, J) ,
\end{equation}
to be substituted into the reduced HJB of Eq.~\ref{reducedHJB}.

\subsection{Adiabatic quadratic approximation and constant slippage}

The logic is the same as with unconstrained rejections.
We freeze reputation and replace the state-dependent reject jump $J(t, q, R)$ by a constant $J \le 0$ defined with self-consistent adiabatic closure.
This yields an inventory-only HJB, which is solved using second-order Hamiltonian expansion and quadratic ansatz for the value function.
Main findings remain the same.
Under constant slippage, $m_n(\delta) \equiv -\theta_n$, we observe significant simplifications:
\begin{itemize}
\item Translational invariance in the shadow price $p$ (hence a shift rule for optimal quotes).
\item Closed-form relations for Hamiltonian derivatives in terms of the myopic optimizer at $p=0$.
\end{itemize}

For fixed $\epsilon$ and $p$, the bucket objective is
\begin{equation}
f_n(\delta,\epsilon; p, J) = \Lambda^n_0 e^{-\kappa_n\delta} \mathcal{G}_n(\delta,\epsilon;p, J) ,
\end{equation}
which is the exponential intensity multiplied by an affine function of $\delta$.
Its interior maximizer is explicit:
\begin{equation}
\delta^*_n(p,\epsilon,J) = p + \bar{\delta}_n(\epsilon, J) = p + \left( \frac{1}{\kappa_n} 
- \frac{J \Phi\big(a_n(\epsilon)\big) + z_n C_n(\epsilon)}{z_n B_n(\epsilon)} \right) .
\end{equation}
Substituting $\delta = \delta^*_n(0, \epsilon, J) =: \bar{\delta}_n(\epsilon, J)$ into the objective yields a one-dimensional maximization problem:
\begin{equation}
\left(\bar{\epsilon}_n(J), \bar{\delta}_n(J)\right) \in \text{arg} \max_{\epsilon \ge 0} f_n\left( \bar{\delta}_n(\epsilon, J), \epsilon, 0, J\right), \qquad
\bar{\delta}_n(J) :=  \bar{\delta}_n(\bar{\epsilon}_n(J), J) .
\end{equation}
There is generally no closed form for $\bar{\epsilon}_n(J)$, but numerical 1D maximization is cheap.

Because $m_n$, $C_n$ and $B_n$ do not depend on $\delta$, the Hamiltonian depends on $p$ only through the combination $u = \delta-p$.
Therefore,
\begin{equation}
H^n(p, J) = e^{-\kappa_n p} H^n(0, J),
\end{equation}
as before.
Moreover, using FOC at the optimizer yields a particularly simple identity
\begin{equation}
H^n_{pp}(0, J) = \Lambda^n_0 e^{-\kappa_n \bar{\delta}_n(J)} z_n \kappa_n B_n(\bar{\epsilon}_n(J)) .
\end{equation}
This is a direct analogue of the earlier formula for unconstrained rejections, with the acceptance probability $\Phi(\tilde{\mu}_n)$
replaced by the protocol acceptance probability $B_n(\bar{\epsilon}_n)$.
With this substitution, optimal quotes are still given by Eq.~\ref{oquotes}, and fair control is a symmetric tolerance
\begin{equation}
\epsilon^*_n = \bar{\epsilon}_n(J) .
\end{equation}

In order to ensure self-consistency for reputation under the protocol, we note that with multiple buckets, a natural closure is the intensity-weighted
average rejection probability at the myopic point $p=0$:
\begin{equation}
\bar{r}(J) = \frac{\sum_{n=1}^N \Lambda^n_0 e^{-\kappa_n \bar{\delta}_n(J)} r_n(J))}{\sum_{n=1}^N \Lambda^n_0 e^{-\kappa_n \bar{\delta}_n(J)}} ,
\qquad r_n(J) := \mathbb{P}(Y < -\bar{\epsilon}_n(J)) = \Phi(a_n(\bar{\epsilon}_n(J))) .
\end{equation}
$J$ is calibrated by solving $\bar{r}(J) = R$, as before.

\begin{figure}[ht]
\centering
\includegraphics[width=0.7\columnwidth]{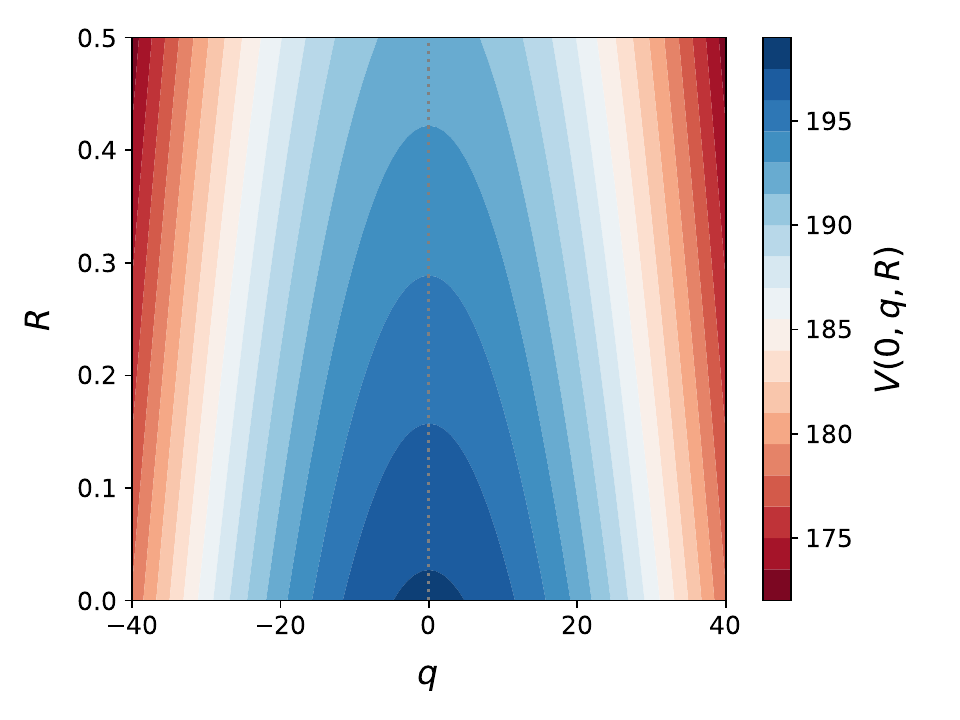}
\caption{
Optimal stationary value function $V(0, q, R)$ for symmetric tolerance with capped slippage protocol and rejection feedback.
Parameters in the text.
}
\label{value_function}
\end{figure}

\subsection{Numerical examples}

Here we consider the same basic parameter set as earlier, with specifics: $\theta_n = 0.1$ bp, $\tau_n \equiv \tau = 0.5$ s, $\rho = 0.01$, $\rho_g = 0.1$.
The dealer controls $(\delta, \epsilon)$.
Figure~\ref{value_function} illustrates the corresponding stationary value function.
The gradient toward reducing inventory and improving reputation is very clear.
Table~\ref{tab:model-comparison} further compares different optimal protocols with and without reputation feedback for an intentionally large latency setting of $\tau = 1$ s and a constant slippage of $\theta_n = 0.1$ bp (except the reference entry with no adverse selection where $\theta_n = 0$).
The results were obtained by numerically solving the corresponding HJB.
We observe that adverse selection significantly harms performance, and the MM responds by widening spreads.
With unconstrained rejections, the utility improves, but clearly at the expense of a high reject rate.
The symmetric protocol results in a materially lower reject ratio and a smaller utility improvement due to rebates.
It is important to note that the optimal threshold level is very reasonable in this case and corresponds to approximately half the top-of-book spread.
Reputation feedback causes both protocols to reduce rejections dramatically.

\begin{table}[h!]
\centering
\begin{tabular}{lccccc}
\toprule
  Model & $\rho_g$ & Spread (bp) & Threshold (bp) & Rejection & Utility\tablefootnote{$V(0, 0, R^*)$.} \\
\midrule
  no rejection, no adverse selection 	& $0$ & $0.69$ & $\infty$ & $0$ & $235.4$ \\
  no rejection 				& $0$ & $0.89$ & $\infty$ & $0$ & $195.1$ \\
  unconstrained 				& $0$ & $0.53$ & $0.25$ & $0.33$ & $201.7$ \\
		    				& $0.05$ & $0.64$ & $0.40$ & $0.22$ & $199.5$ \\
		    				& $0.10$ & $0.72$ & $0.54$ & $0.15$ & $198.3$ \\
		    				& $0.15$ & $0.76$ & $0.67$ & $0.10$ & $197.4$ \\
		    				& $0.20$ & $0.80$ & $0.79$ & $0.06$ & $196.9$ \\
  symmetric       				& $0$ & $0.71$ & $0.43$ & $0.17$ & $199.4$ \\
		    				& $0.05$ & $0.75$ & $0.49$ & $0.13$ & $198.2$ \\
		    				& $0.10$ & $0.78$ & $0.55$ & $0.09$ & $197.4$ \\
		    				& $0.15$ & $0.81$ & $0.61$ & $0.07$ & $196.9$ \\
		    				& $0.20$ & $0.83$ & $0.68$ & $0.05$ & $196.5$ \\
\bottomrule
\end{tabular}
\caption{Comparison of optimal trade acceptance/rejection protocols. 
``Threshold'' denotes the optimal one-sided rejection cutoff in the unconstrained model and the optimal symmetric tolerance $\epsilon$ in the fair protocol.
Parameters are given in the text.}
\label{tab:model-comparison}
\end{table}

\subsubsection*{Simulation}

Finally, we provide a Monte Carlo simulation of market making with adverse selection and reputation feedback under a stationary optimal policy.
We consider constant slippage here and assume that the mechanism is driven by endogenous adverse quote selection.
The simulator steps in units of latency $\tau$ and models a single step as $\Delta S \sim \mathcal{N}(0, \sigma^2 \tau)$.
RFQs are then chosen with tilted intensities 
\begin{equation}
\lambda^{n,i}(\Delta S) = \lambda^{n,i}_\text{base} \exp(-\alpha_n Y^i) ,
\end{equation} 
where $Y^i = \pm \Delta S$ and
$\alpha_n = \theta_n / (\sigma^2 \tau)$ so that $Y | \text{RFQ} \sim \mathcal{N}(-\theta_n, \sigma^2 \tau)$.
In order to keep the unconditional RFQ rate unchanged, we include the intensity rescaling 
\begin{equation}
\Lambda^n_{0, \text{adj}} = \Lambda^n_0 \exp\left(-\frac{1}{2}\alpha_n^2 \sigma^2\tau\right) .
\end{equation}
This rescaling exactly offsets the Gaussian exponential tilt because, under the untilted law, $\mathbb{E}\left[\exp\left(-\alpha_n Y^i\right)\right] = \exp\left(\tfrac{1}{2} \alpha^2_n \sigma^2 \tau\right)$.
Hence the unconditional RFQ arrival rate remains equal to the original baseline level.
The algorithm uses exact stationary controls obtained by numerically solving the corresponding HJB
and supports both the unrestricted and fair protocols.

\begin{figure}[ht]
\centering
\includegraphics[width=0.7\columnwidth]{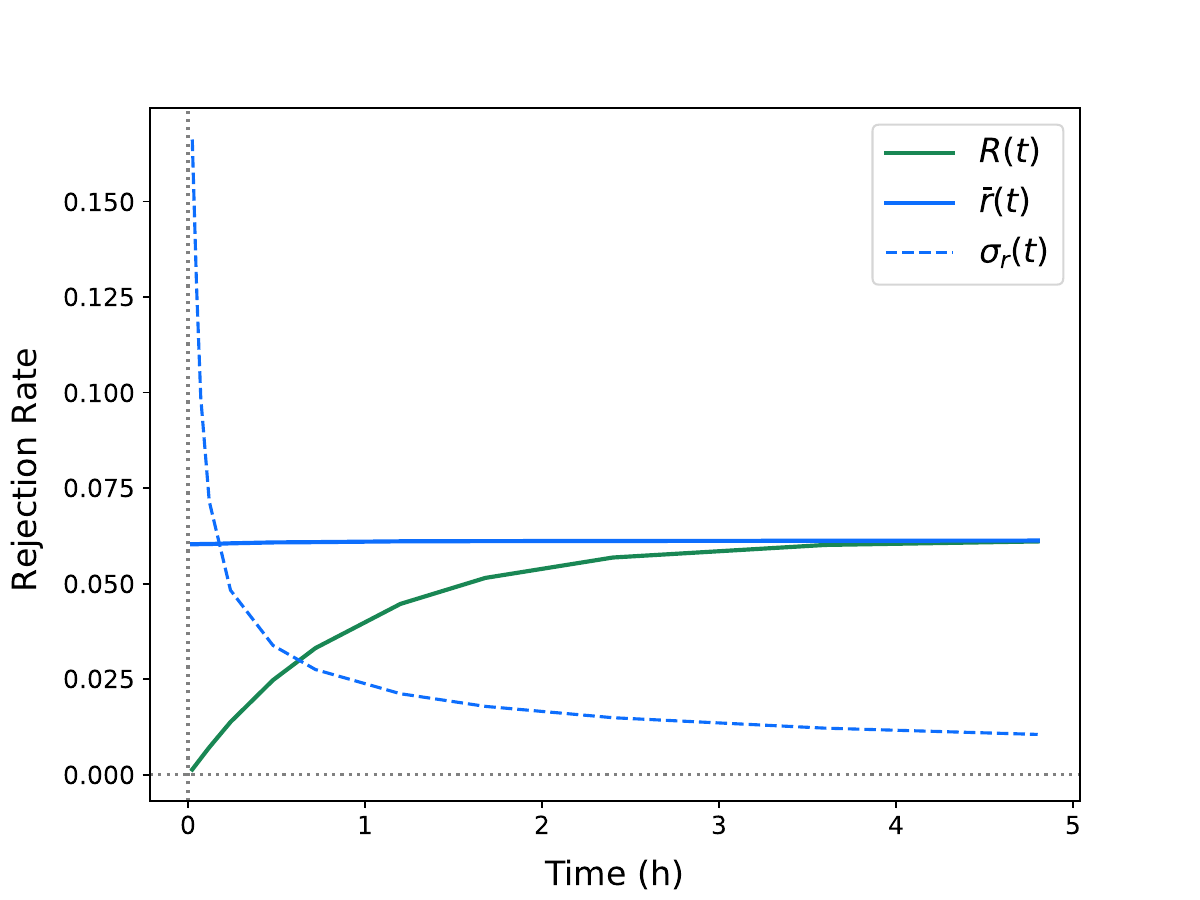}
\caption{
Relaxation of the simulated rejection score (rejection-rate EMA) to the stationary state.
The mean rejection rate is optimal from the start, with the standard deviation decreasing as the stationary state is approached.
Parameters in the text.
}
\label{sim_rejections}
\end{figure}

\begin{figure}[ht]
\centering
\includegraphics[width=0.7\columnwidth]{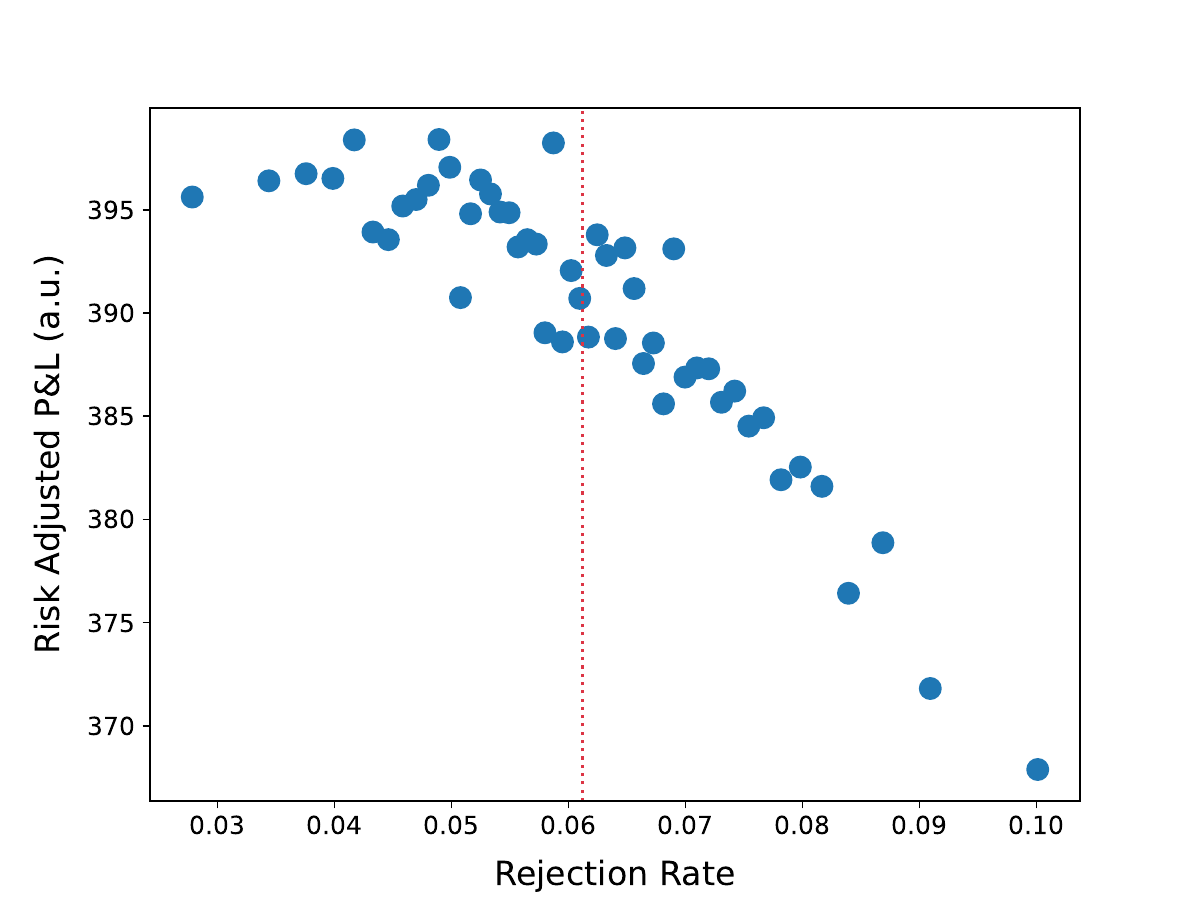}
\caption{
Inventory risk adjusted P\&L (optimization objective) as a function of the realized rejection rate at $T = 2.4$ hours.
Circles are averages of a thousand consecutive sorted realizations.
The dashed line corresponds to the stationary state.
Parameters in the text.
}
\label{sim_utility}
\end{figure}

Simulation was run for the fair acceptance/rejection protocol with latency $\tau_n = 0.5$ s, constant slippage $\theta_n = 0.1$ bp and reputation parameters $\rho = \rho_g = 0.01$, a total of $5\cdot 10^4$ trajectories.
Optimal controls from the HJB solution were passed to the simulator with initial conditions $q = 0$, $R = 0$.
Figure~\ref{sim_rejections} illustrates relaxation of the reputation variable to the stationary state as the dealer maintains the optimal rejection rate from the start.
Initially, with a low rejection score (small $R$), the dealer is more flexible in their decision, as evidenced by the high standard deviation.
Figure~\ref{sim_utility} demonstrates how the value function (inventory risk adjusted P\&L) depends on the realized rejection rate.
This kind of plot can be verified empirically.

\section{Concluding remarks}

FX dealers respond to latency risk with spread widening and rejection protocols.
So far, these protocols have been a matter of design rather than optimization, with fairness and transparency being in the spotlight \citep{gfxc2024}.
In this paper, we provide a tractable framework that builds on the Avellaneda-Stoikov tradition \citeyearpar{avellaneda2008} and treats the rejection decision --
subject to fairness or protocol constraints -- as an optimizable state-dependent control alongside quoting.
Optimal thresholds are found to be conservative, even in the presence of toxic flow, although the actual values depend on market conditions.
We have also introduced the so-called rejection-score feedback mechanism to model how frequent rejections can reduce subsequent client engagement.
Rejection-score-aware control makes the dealer even less inclined to reject in order to protect the flow.
An adiabatic quadratic approximation with self-consistent closure was suggested as a simple, practical way to calculate optimal controls.

\section*{Acknowledgments}
The author is grateful to Eric Mathew John (HSBC) for fruitful discussions and to Richard Anthony (HSBC) for support throughout the project and valuable comments.
The views expressed are those of the author and do not necessarily reflect the views or practices at HSBC.

\end{document}